\documentclass[conference]{IEEEtran}

\usepackage{booktabs} 
\usepackage{enumerate}
\usepackage{adjustbox}
\usepackage{graphicx}
\usepackage{amsmath}
\usepackage{multirow}
\usepackage{footnote}
\usepackage{threeparttable}
\usepackage{amsmath}
\usepackage{textcomp}
\usepackage{balance}
\usepackage{hyperref}
\usepackage{todonotes}
\usepackage{array}
\usepackage{fontawesome5}
\usepackage{cite}
\usepackage{listings}
\usepackage{tabularx}

\usepackage{tikz}

\usepackage{colortbl}
            
\usepackage{amssymb}
\usepackage{pifont}

\definecolor{lightgray}{gray}{0.9}

\usepackage[square, comma, numbers]{natbib}

\usepackage{graphicx}

\usepackage{algpseudocode}

\usepackage[caption=false]{subfig}
\usepackage{listings}

\begin{document}

\title{JEDI-linear: Fast and Efficient Graph Neural Networks for Jet Tagging on FPGAs}

\author{
    \IEEEauthorblockN{
        Zhiqiang Que\IEEEauthorrefmark{4}\IEEEauthorrefmark{1},
        Chang Sun\IEEEauthorrefmark{4}\IEEEauthorrefmark{2},
        Sudarshan Paramesvaran\IEEEauthorrefmark{3},
        Emyr Clement\IEEEauthorrefmark{3},
        Katerina Karakoulaki\IEEEauthorrefmark{3}, \\
        Christopher Brown\IEEEauthorrefmark{5},
        Lauri Laatu\IEEEauthorrefmark{1},
        Arianna Cox\IEEEauthorrefmark{1},
        Alexander Tapper\IEEEauthorrefmark{1},
        Wayne Luk\IEEEauthorrefmark{1},
        Maria Spiropulu\IEEEauthorrefmark{2}
    }

    \vspace{0.2cm}

    \IEEEauthorblockA{
        \IEEEauthorrefmark{1}
        Imperial College London, UK,
        \{z.que, l.laatu, a.cox24, a.tapper, w.luk\}@imperial.ac.uk
    }

    \IEEEauthorblockA{
        \IEEEauthorrefmark{2}
        California Institute of Technology, Pasadena, CA, USA,
        \{chsun, smaria\}@caltech.edu
    }

    \IEEEauthorblockA{
        \IEEEauthorrefmark{3}
        University of Bristol, UK,
        \{sudarshan.paramesvaran, emyr.clement, k.karakoulaki\}@bristol.ac.uk
    }

    \IEEEauthorblockA{
        \IEEEauthorrefmark{5}
        European Organization for Nuclear Research (CERN), Switzerland,
        christopher.edward.brown@cern.ch,
    }

    \vspace{-0.7cm}
}

\maketitle

\begingroup\renewcommand\thefootnote{\IEEEauthorrefmark{4}}
\footnotetext{Equal contribution.}
\endgroup

\begin{abstract}

   Graph Neural Networks (GNNs), particularly Interaction Networks (INs), have shown exceptional performance for jet tagging at the CERN High-Luminosity Large Hadron Collider (HL-LHC). However, their computational complexity and irregular memory access patterns pose significant challenges for deployment on FPGAs in hardware trigger systems, where strict latency and resource constraints apply. In this work, we propose JEDI-linear, a novel GNN architecture with linear computational complexity that eliminates explicit pairwise interactions by leveraging shared transformations and global aggregation. To further enhance hardware efficiency, we introduce fine-grained quantization-aware training with per-parameter bitwidth optimization and employ multiplier-free multiply-accumulate operations via distributed arithmetic. Evaluation results show that our FPGA-based JEDI-linear achieves 3.7 to 11.5 times lower latency, up to 150 times lower initiation interval, and up to 6.2 times lower LUT usage compared to state-of-the-art GNN designs while also delivering higher model accuracy and eliminating the need for DSP blocks entirely. This is the first interaction-based GNN to achieve less than 60~ns latency and currently meets the requirements for use in the HL-LHC CMS Level-1 trigger system. This work advances the next-generation trigger systems by enabling accurate, scalable, and resource-efficient GNN inference in real-time environments. Our open-sourced templates will further support reproducibility and broader adoption across scientific applications.

\end{abstract}

\section{Introduction}

Modern high energy physics (HEP) experiments, such as those at the CERN Large Hadron Collider (LHC), generate massive volumes of high-dimensional data, amounting to hundreds of terabytes per second, from particle collisions happening every 25~ns~\cite{coelho2021automatic, duarte2018fast}. It is neither technically feasible nor scientifically useful to store all collision events, as the vast majority do not contain interesting or rare physics. To address this, a two level trigger system is designed to rapidly filter and keep only the most interesting events for further analysis. The first stage, known as the Level-1 Trigger (L1T), operates under strict real-time constraints, making decisions within a few microseconds. It is composed of hundreds of FPGAs that perform low-latency computations on the incoming data. Jet tagging is a crucial task in this process, as it enables the identification of jets, which are collimated sprays of particles produced in high-energy collisions. Accurate jet identification is essential for distinguishing between different types of collision events and making effective trigger decisions.

At L1T, a jet is represented by a sequence of particles that are clustered in a spatial region. As such data exhibit complex relational patterns, they are ideal candidates for processing with Graph Neural Networks (GNNs), which have shown great promise in modeling such interactions~\cite{moreno2020jedi, que2024ll}. By learning over graph-structured inputs, GNNs can capture both local and global particle correlations, enabling more effective event classification and jet tagging. However, despite their algorithmic strengths, GNNs remain difficult to deploy in real-time systems such as the L1T at the LHC, which must operate under extreme constraints: as one stage in the L1T system, the jet's class must be inferred in a few hundred nanoseconds, and no more than one Super Logic Region (SLR) of a single FPGA~\cite{summers2024twepp, summers2023p2upgrade}, or even less depending on the other algorithms running in parallel, can be used for the jet tagging algorithm. This requires the GNN to be highly efficient in terms of both latency and hardware resource usage while still maintaining high model accuracy.

Recent efforts have sought to adapt GNN architectures on FPGAs to meet the strict latency constraints of particle physics applications. For example, LL-GNN~\cite{que2024ll} demonstrates that a GNN-based algorithm, JEDI-net~\cite{moreno2020jedi} can achieve inference within 1 microsecond using highly optimized FPGA implementations. Similarly, the ultrafast JEDI-net introduced in~\cite{odagiu2024ultrafast} employs 8-bit quantization to reduce model size while maintaining high accuracy with low latency, achieving superior accuracy than other architectures such as DeepSets~\cite{zaheer2017deep}. However, its high on-chip resource usage makes it impractical to deploy on real-world FPGA systems.

To address these challenges, we propose JEDI-linear, a novel variant of the GNN-based JEDI-net algorithm that is efficient and scalable while preserving its strong modeling capacity. In particular, we show that by requiring the edge-wise interaction function to be an affine transformation, we can eliminate the explicit pairwise edge computations and instead aggregate the particle features globally, which can be used to reduce the overall complexity from $\mathcal{O}(N_O^2)$ to $\mathcal{O}(N_O)$, where $N_O$ is the maximum number of particles in a jet.

To maximize hardware efficiency on FPGAs, we apply fine-grained quantization-aware training with per-parameter bitwidth optimization. In contrast to uniform quantization schemes~\cite{odagiu2024ultrafast, coelho2021automatic}, our approach allows each parameter to adopt a bitwidth tailored to its impact on model performance. This is particularly well-suited for fully unrolled designs with strict latency requirements, in which each operation is mapped to dedicated hardware unit and reuse of these units for arithmetic operations is avoided. As a result, we can assign custom bitwidths to each operation, enabling fine-grained trade-offs between accuracy and resource usage. This bit-level control fully exploits the flexibility of reconfigurable logic on FPGAs, 
improving both latency and hardware efficiency.

Moreover, we integrate Distributed Arithmetic (DA) to further optimize the implementation of multiply-accumulate (MAC) operations. DA replaces conventional multipliers with adder graphs, leveraging FPGA-friendly structures to implement high-throughput, low-resource arithmetic units. When used in conjunction with our quantized model and unrolled architecture, DA allows us to significantly reduce the number of LUTs used and eliminate all DSP block usage, further increasing the feasibility of real-time deployment.

To the best of our knowledge, this is the first GNN for jet tagging that meets both the resource and latency requirements of a real-world L1T system at CERN. This work paves the way for next-generation hardware-based trigger systems by enabling powerful, low-latency algorithms to process massive LHC data streams efficiently.

We make the following contributions in this paper:
\begin{itemize}
   \item A novel linear-complexity interaction-based GNN architecture (JEDI-linear), which removes explicit pairwise edge computations while preserving global context aggregation. We also apply fine-grained quantization-aware training with per-parameter quantization and implement multiplier-free MAC operations using distributed arithmetic. These algorithm and hardware optimization strategies significantly reduce hardware usage and latency while maintaining model accuracy, enabling scalable, low-latency inference for real-time high energy physics applications.

   \item A scalable, low-latency JEDI-linear template that enables the generation of efficient FPGA implementations using high-level synthesis (HLS) as well as Verilog, with a focus on minimizing latency and resource usage. We open-source the JEDI-linear templates and the code\footnote{https://github.com/calad0i/JEDI-linear} to generate the proposed hardware designs to support reproducibility and benefit the wider research community.

   \item A comprehensive evaluation of the proposed approach and hardware implementation.
\end{itemize}

Although this work focuses on jet tagging in high-energy physics, the underlying principles and optimization techniques, such as linearized processing, mixed-precision quantization, distributed arithmetic, and fully unrolled hardware design, are broadly applicable. They can be adapted for trustworthy DNNs~\cite{que2025trustworthy} as well as low-latency variational autoencoders (VAEs)~\cite{que2024low}, transformers~\cite{wojcicki2022accelerating}, and LLMs~\cite{que2025asicon}.

\section{Background and Related Work} \label{sec:background}

\begin{figure}
   \begin{center}
      \includegraphics[width=0.7\linewidth]{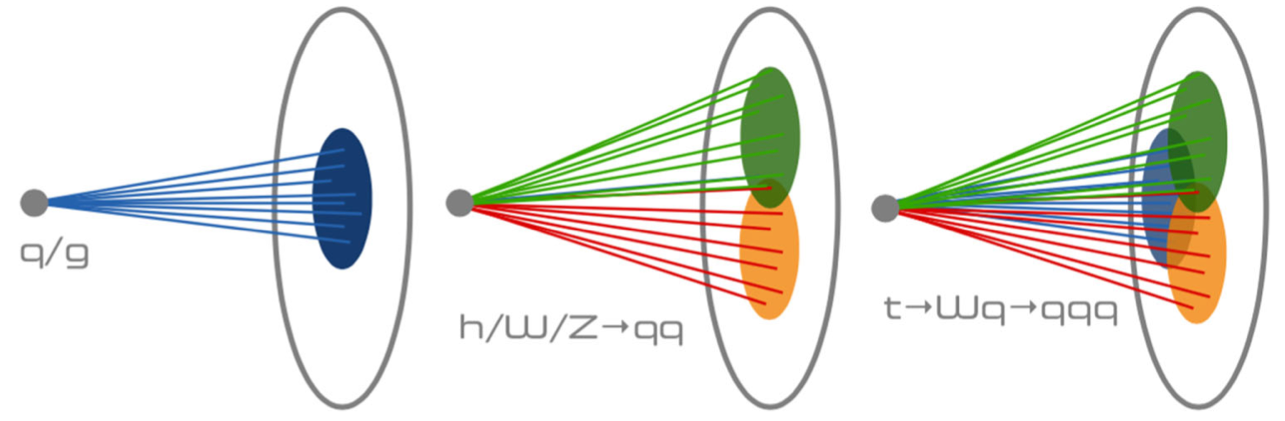}
   \end{center}
   \vspace{-0.3cm}
   \caption{Schematic representation of various jet types in particle physics~\cite{moreno2020jedi}. These differences in jet topology are exploited by jet tagging algorithms.}
   \label{fig:jets}
\end{figure}

\subsection{Jet Tagging in Particle Physics}

In high-energy particle collisions at the CERN LHC, quarks and gluons cannot exist freely due to color confinement and instead hadronize into collimated sprays of particles known as jets~(Fig.~\ref{fig:jets}). Jet tagging identifies the jet's origin, such as b-quarks, c-quarks, gluons, or heavy particles (W/Z, Higgs, top), and is vital for precise event classification and new physics searches.
While early methods relied on handcrafted features and cut-based techniques, modern approaches increasingly use machine learning. Recent developments include the use of GNNs~\cite{moreno2020jedi, qu2020jet} which treat jets as graphs of constituent particles and learn representations from low-level detector inputs. These models offer improved tagging performance and are particularly promising for real-time applications in L1T systems when combined with fast and efficient hardware like FPGAs. Transformer-based models such as ParT~\cite{qu2022particle, wang2024interpreting} also demonstrate good model accuracy but are typically too large for FPGA deployment in L1T. Instead, they are more suitable for offline uses, or potentially the high-level trigger systems after L1T, where latency budgets are less stringent. Therefore, this work focuses on optimizing GNN-based architectures for L1T applications where sub-100~ns latency and sub-10~ns initiation intervals are required.

\subsection{Graph Neural Networks on FPGAs for Jet Tagging}
GNNs can capture complex substructure and exploit low-level detector information, leading to state-of-the-art performance~\cite{moreno2020jedi, qu2020jet} in distinguishing different jet types. However, deploying GNNs in real-time environments poses significant challenges due to the stringent low latency and resource constraints.

Several studies have explored the use of Multi-Layer Perceptrons (MLP) networks on FPGAs for jet tagging~\cite{duarte2018fast, coelho2021automatic}, but the achieved model accuracy is relatively low. Recent efforts~\cite{que2022reconf, que2022opt, zhang2023particlenet, que2024ll, odagiu2024ultrafast} focus on co-designing GNN architectures with hardware-aware optimization techniques, such as quantization, pruning, and efficient graph representations, to meet the latency and resource requirements of FPGA deployment at L1T while preserving model accuracy. Ref.~\cite{que2022reconf} presents the first FPGA-based implementation of the GNN-based JEDI-net~\cite{moreno2020jedi} for jet tagging, achieving sub-microsecond initiation intervals using a custom matrix-matrix multiplication (MMM) that exploits sparsity and binary features to reduce unnecessary multiplications and hardware usage.
Ref.~\cite{zhang2023particlenet} presents another GNN-based algorithm, ParticleNet, on FPGAs. However, the achieved latency (15 ms) is too high for L1T applications. Ref.~\cite{que2022opt} introduces an outer-product-based MMM approach and a two-level parallelism strategy, reducing the latency to 1.57~$\mu$s and 10.66~$\mu$s for the 30-particle and 50-particle JEDI-net models, respectively. To further reduce latency, LL-GNN~\cite{que2024ll} adopts task-level parallelism, sublayer fusion, and latency-aware algorithm-hardware co-design, resulting in a sub-microsecond latency. However, it still incurs a minimum initiation interval of 150 ns, which is too large. It also consumes large hardware resources, with over 8,700 DSPs. Similarly, \cite{odagiu2024ultrafast} applies quantization-aware training and utilizes a uniform 8-bit fixed-point representation, leading to a latency of around 150 ns but still requiring large hardware resource consumption, e.g., it requires over 5,000 DSPs and 1,388k LUTs for a 16-particle JEDI-net model.

Despite these advancements, these prior FPGA-based GNN designs are still impractical to deploy in the L1T system. In particular, the explicit all-to-all edge-level computations require significant amount of hardware resources to perform, resulting in large latencies as well as initiation intervals, which hinders design scalability. In contrast, this work proposes a novel GNN architecture, JEDI-linear, which eliminates the explicit edge-level operations entirely and achieves major reductions in resource consumption. Combined with the fine-grained quantization-aware training and distributed arithmetic, JEDI-linear achieves significant reductions in latency and resource usage while maintaining high model accuracy, making GNNs practical for real-time jet tagging applications in the L1T system.

\section{Design and Optimization} \label{sec:design}

\subsection{Global Information Gathering}

\begin{figure}
   \begin{center}
      \includegraphics[width=1.0\linewidth]{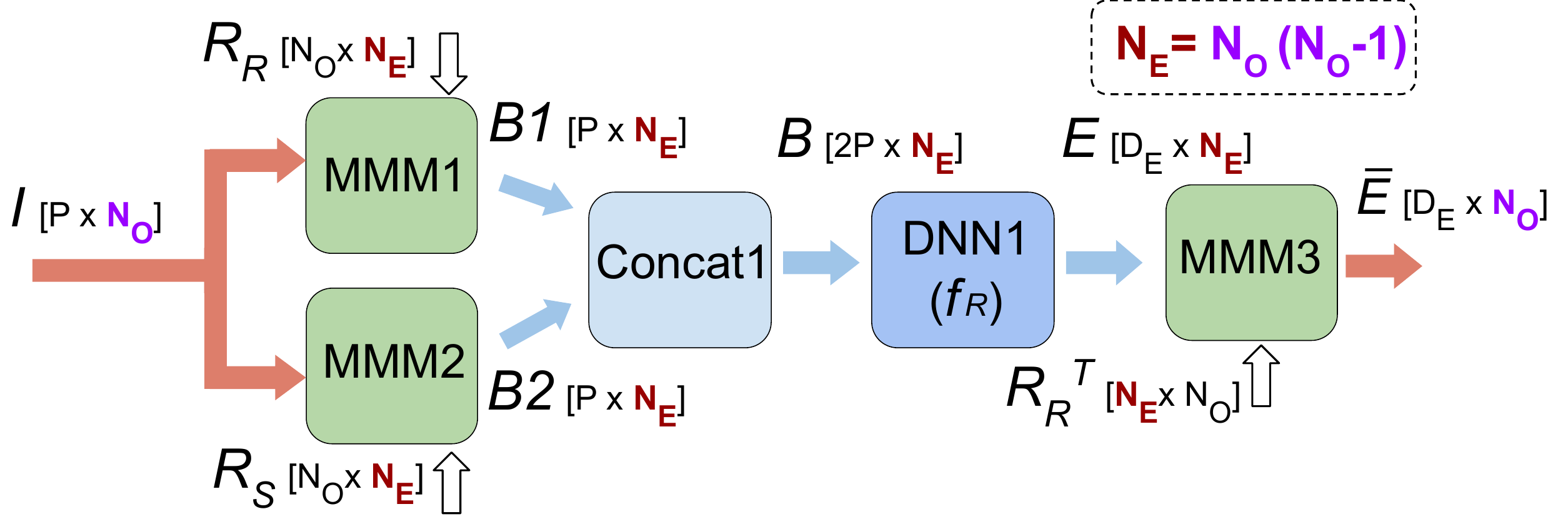}
   \end{center}
   \vspace{-0.3cm}
   \caption{Conventional interaction information gathering in JEDI-net. $R_R$ and $R_S$ are receiving and sending matrices; $N_O$ and $N_E$ are the numbers of particles and edges. $B1, B2, B, E, E, \bar{E}$ are all intermediate variables. }
   \label{fig:interaction_old}
\end{figure}

The original JEDI-net models a jet as a densely connected, directed graph without self loops. Each input, totaling $N_O$ particles, is regarded as a node, and a directional edge is defined between all particles. For all edges, a learnable function $f_R$ processes the concatenated features of the source and destination particles, yielding an edge interaction embedding, as shown in Fig.~\ref{fig:interaction_old}. The interaction embeddings are then gathered for each particle by a summation on the destination particles, resulting in a new feature representation for each particle that incorporates information from all other particles in the jet. Denoting $P$ as the number of features per particle, the input can be represented in a 2D array of $I \in \mathbb{R}^{N_O \times P}$.
We denote $i,j\in [1, N_O]$ where $i\ne j$ to be the indices of the source and destination particles.
In addition, we define the concatenation of source and destination particle feature vectors $I_i$ and $I_j$ as $I_i \| I_j \in \mathbb{R}^{2P}$.
The interaction embedding can be obtained as follows:
\begin{align}
   \label{eq:jedi_original}
   \begin{split}
      B_{ij}    & = I_i \| I_j \in \mathbb{R}^{2P}
      \\
      E_{ij}    & = f_R(B_{ij}) \in \mathbb{R}^{D_E}     \\
      \bar{E}_i & = \sum_{j\ne i} E_{ij} \in \mathbb{R}^{D_E}
   \end{split}
\end{align}
Here, $f_R: \mathbb{R}^{2P} \rightarrow \mathbb{R}^{D_E}$ is a learnable function, which is implemented as a neural network in the original work~\cite{moreno2020jedi}. $D_E$ denotes the edge latent dimension. 
Since $f_R$ is applied to every pair of particles, the overall computational complexity of this step is $\mathcal{O}(N_O^2 \cdot C_{f_R})$, where $C_{f_R}$ denotes the cost of evaluating $f_R$ once. This quadratic scaling becomes a significant bottleneck when $N_O$ is even moderately large. For instance, $N_O = 64$ would lead to over 4,000 interactions, making it infeasible for real-time applications. As a result, the pairwise interaction step dominates the inference cost and poses significant challenges for low-latency applications and hardware-constrained deployment environments, such as real-time triggering on FPGA-based systems.

\subsection{The Proposed Global Information Gathering}
Several previous studies have attempted to address the computational bottleneck inherent in the JEDI-net formulation when deployed on FPGAs. In~\cite{que2022reconf}, a custom MMM (matrix-matrix multiplication) module is proposed for GNN feature transformation. By leveraging the sparse structures and binary features within the adjacency matrices $R_R$ and $R_S$, it can pick out elements and suppress unnecessary computations instead of conducting real multiplication operations.
In~\cite{que2022opt}, an outer-product-based MMM approach is introduced using a coarse-grained pipeline to improve throughput. More recently,~\cite{que2024ll} proposes a fused Edge-Node unit that integrates edge and node computations into a single hardware module. However, all of these designs still involve explicit edge-level computations, as shown in Fig.~\ref{fig:interaction_old}, resulting in poor scalability and significant hardware resource consumption.

\begin{figure}
   \begin{center}
      \includegraphics[width=0.90\linewidth]{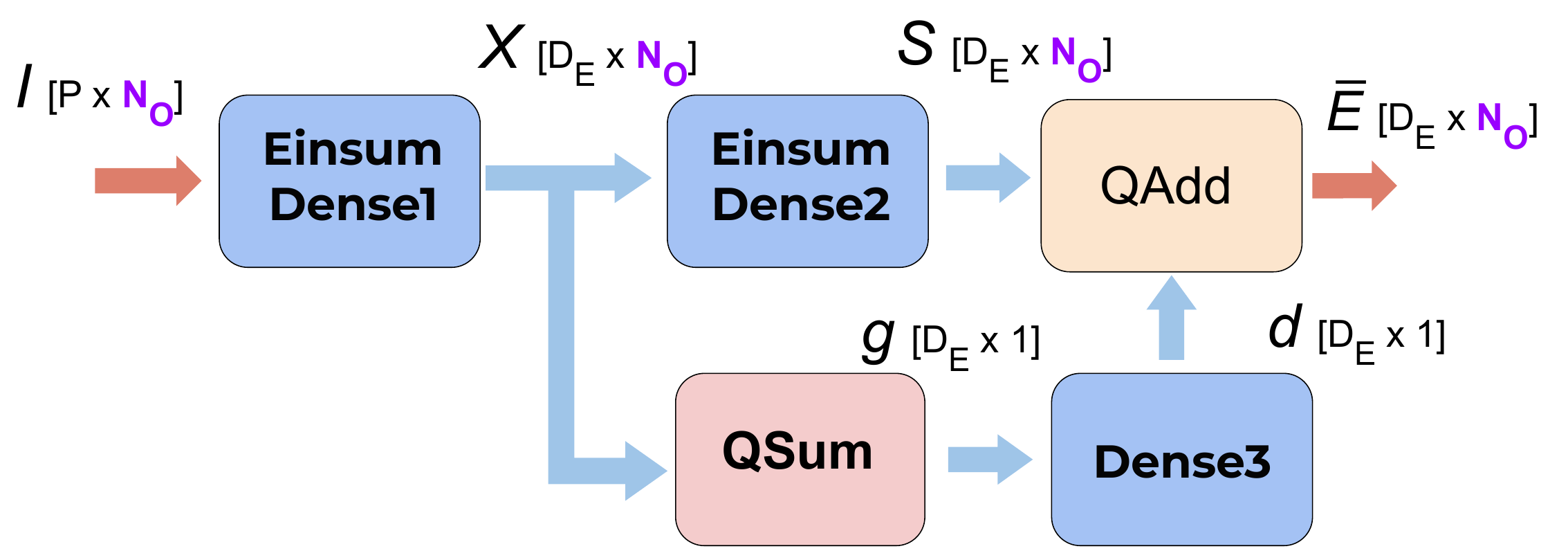}
   \end{center}
   \vspace{-0.3cm}
   \caption{The proposed interaction information gathering for JEDI-net. Einsum denotes Einstein summation.}
   \label{fig:interaction_new}
\end{figure}

In this work, we introduce JEDI-linear, a variant of JEDI-net for which the computation cost scales linearly with the number of particles in a jet, $N_O$, to address the computational bottleneck inherent in the original JEDI-net formulation. To achieve this, we require $f_R$ to be an \textit{affine} transformation, or a single dense layer:
\begin{equation}
   \label{eq:affine}
   f_R(B_{ij})_{l} = W \left(I_i \| I_j\right) + C = W_1 I_i + W_2 I_j + C
\end{equation}
where $W_1, W_2 \in \mathbb{R}^{D_E \times P}$ are the learned weights for the source and destination particles, respectively, and $C \in \mathbb{R}^{D_E}$ is a bias term. With this assumption, we can rewrite the interaction embedding in~\eqref{eq:jedi_original} as follows:
\begin{align}
   \label{eq:jedi_original_elementwise}
   \begin{split}
      \bar{E}_i &= \sum_{j\ne i} f_R(I_i \| I_j) \\
      &= \sum_{j\ne i}\left(W_1 I_i + W_2 I_j + C\right) \\
      &= W_2 \sum_{j}I_j - W_2I_i + \left(N_O-1\right)\left(W_1 I_i + C\right) 
   \end{split} 
\end{align}
Assuming $N_O>>1$, we scale everything by $1/N_O$ to avoid numerical instability and ignore terms of the order of $1/N_O$ or smaller, we can rewrite~\eqref{eq:jedi_original_elementwise} as:
\begin{align}
   \label{eq:jedi_linear}
   \begin{split}
      \bar{E}_i' &= \frac1{N_O} \bar{E}_i \\
      &= W_2\frac1{N_O} \sum_{j}I_j - \frac{W_2I_i}{N_O} + \frac{N_O-1}{N_O}\left(W_1 I_i + C\right) \\
      &\approx W_2 \frac1{N_O} \sum_{j}I_j + W_1 I_i + C 
   \end{split}
\end{align}
As shown in \eqref{eq:jedi_linear}, the interaction embedding $\bar{E}_i'$ can now be computed with linear complexity with respect to $N_O$. The first term can be regarded as a global context vector summarizing the features of all particles, and the second term is the transformation of the features of all individual particles. In implementation, we can compute the global context vector by a global average pooling over the first dimension followed by a dense layer, and the second term is implemented by another dense layer applied along the first axis. The bias term $C$ is absorbed into any of the two dense layers. Results of the two dense layers are then added together to obtain the final interaction-aware feature representation for each particle. We show the whole global information gathering in Fig~\ref{fig:interaction_new}.

With the reduced computational complexity, we find that the network is highly scalable and efficient, especially for jets with a large number of particles. While JEDI-linear does not explicitly model every pairwise interaction, it preserves the inductive bias that each particle's representation should be influenced by the collective behavior of the jet.

\begin{figure}
   \begin{center}
      \includegraphics[width=1.0\linewidth]{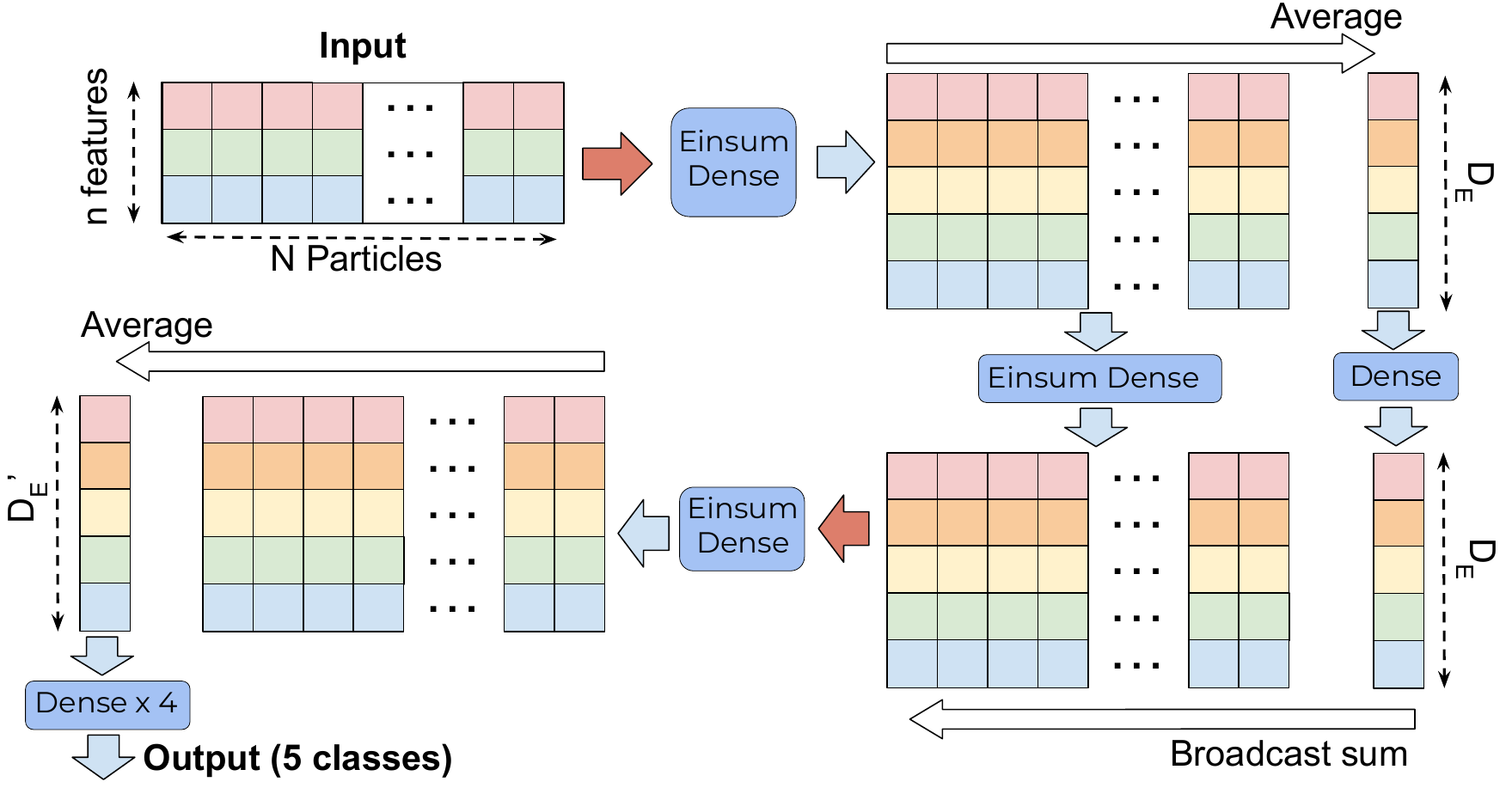}
   \end{center}
   \caption{The architecture of the JEDI-linear model. The input projection layer projects the input features into a latent space, and the global information gathering layer aggregates the latent embeddings across all particles to produce a context vector. The context vector is then transformed and broadcast back to each particle, resulting in an interaction-aware feature representation for each particle.}
   \label{fig:arch}
\end{figure}

The final neural architecture of JEDI-linear is shown in Fig.~\ref{fig:arch}.
The model begins with an input projection layer that transforms each particle's feature vector into a latent representation. This is followed by a global information gathering stage, where the embeddings from all particles are aggregated via average pooling to form a global context vector. This vector captures the overall structure of the jet and is broadcast back to each particle representation. The combined features are then processed by a second shared dense layer, enabling global interaction. Finally, the updated particle features are aggregated again using average pooling and passed through a multi-layer perceptron (MLP) classification head to produce logits for five jet classes.

\subsection{Mixed Quantization and Pruning}\label{sec:HGQ}
To enable efficient deployment of JEDI-linear on resource-constrained hardware, we present a fine-grained mixed-precision quantization scheme that is co-designed with the neural network architecture as well as the unrolled hardware architecture (Section~\ref{sec:hw_arch}). Leveraging the High Granularity Quantization (HGQ) framework~\cite{sun2024gradient}, our approach assigns bitwidths at the level of individual weights during training, guided by their impact on both predictive accuracy and hardware resource usage.
This is achieved via a differentiable surrogate gradient formulation that adjusts bitwidths during training, targeting a trade-off between predictive loss and a differentiable, hardware-aware resource estimator called Effective Bit Operations (EBOPs) incorporated into the loss function.

Unlike post-training quantization or other quantization-aware training workflows that treat hardware constraints outside the training loop, our method of fine-grained quantization-aware training tightly couples model performance and compression. Because each JEDI-linear operation maps to a dedicated hardware block (More details in Sec.~\ref{sec:hw_arch}), per-parameter bitwidth tuning yields immediate and fine-grained reductions in area and latency. Furthermore, we unify quantization and unstructured pruning into a single objective: parameters whose bitwidths are driven to zero are automatically pruned during training. This enables a single training trajectory to explore the Pareto frontier of accuracy versus hardware cost, unlike multi-stage pipelines or iterative sweeps (e.g., AutoQKeras~\cite{coelho2021automatic}). Applied to JEDI-linear, this approach leads to an extremely compact and hardware-efficient design without manual bitwidth tuning or pruning schedules.

\subsection{Distributed Arithmetic for JEDI-linear}\label{sec:DA}

To further reduce latency and resource usage in JEDI-linear's FPGA designs, we adopt a multiplier-free design approach based on Distributed Arithmetic (DA), and integrate it into our end-to-end co-design workflow using the \texttt{da4ml} framework~\cite{sun2025da4ml}. Unlike conventional designs that rely on fixed-point multipliers for constant matrix-vector multiplications (CMVMs), DA decomposes these operations into a static graph of shift-addition/subtraction operations, which can be implemented using highly efficient LUT-logic with fast carriers. Common subexpression elimination is applied to each CMVM operation to reduce the redundant adder/subtractor logic required at multiple places, further improving the overall resource efficiency.

While \texttt{da4ml} was originally designed as an operator-level optimizer, we extend its symbolic tracing capabilities to support the more complex computation patterns found in JEDI-linear and facilitate automatic hardware generation. In particular, we introduce support for global average pooling, broadcast operations, and the underlying Einsum dense layers used to implement the networks. This enables end-to-end tracing of the JEDI-linear computation graph and automatic generation of pipelined DA-based hardware modules. Our enhancements allow \texttt{da4ml} to restructure all CMVM operators in the network, along with other required operations, into the optimized adder graphs. As a result, all MAC operations in JEDI-linear are implemented without DSP usage, while maintaining high operating frequencies and low latency. Our approach enables a fine-grained exploration of the Pareto frontier between resource usage, accuracy, and latency, and provides a pathway toward rapid deployment of learned models in high-throughput real-time environments like the LHC. Moreover, the ability to directly generate synthesizable Verilog from traced models allows for rapid prototyping and hardware validation, which we leverage throughout our design workflow.

\section{Implementation} \label{sec:Implementation}

\subsection{System Overview}

\begin{figure}
   \begin{center}
      \includegraphics[width=0.95\linewidth]{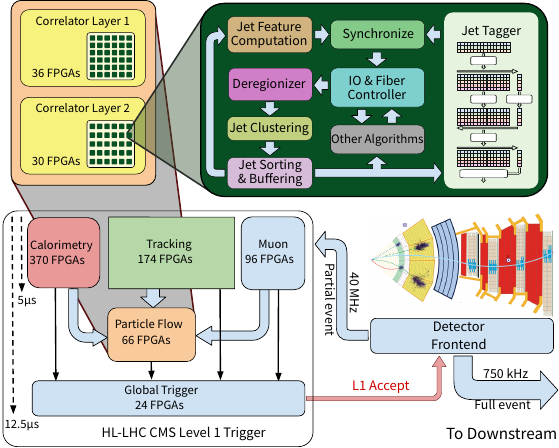}
   \end{center}
   \caption{A sketch of the CMS Level-1 Trigger system in the HL-LHC, adapted from~\cite{summers2024twepp, cms-tdr-021, que2024ll}. The JEDI-linear model proposed in this work targets the FPGAs in the Correlator Layer 2.}
   \label{fig:system}
\end{figure}

The proposed JEDI-linear targets the Correlator Trigger Layer 2 (CTL2) of the CMS Level-1 Trigger system~\cite{summers2024twepp,cms-tdr-021}, as shown in Fig.~\ref{fig:system}. The Level-1 Trigger is a low-latency real-time decision-making system in the CMS detector at the CERN HL-LHC, composed of multiple layers and hundreds of FPGAs organized by subsystem (Calorimetry, Tracking, Muon) and functionality (Correlator, Global Trigger, etc.).
Within the Correlator system, CTL2 comprises 30 VU13P FPGAs, which are organized into 5 slices of 6 nodes each. Each node operates in a round-robin fashion to process incoming events, effectively making 5 FPGAs available for algorithm deployment at any given time. These FPGAs must host both the jet clustering and jet tagging algorithms, making resource and latency efficiency critical for integration.
Within CTL2, the JEDI-linear unit is placed downstream of the jet preprocessing logic, which performs operations such as clustering, sorting, and buffering.
It operates in a pipelined manner, with an initiation interval of 1 clock cycle and a target frequency of over 300 MHz. It is optimized to handle multiple jets per event by processing each jet independently and concurrently. The produced jet tag is then synchronized with the jet features computed in other parallel logic. The synchronized outputs are then passed to subsequent logic for global decision-making. Importantly, the algorithm is required to have deterministic latency for synchronizing with the other tasks.

\subsection{Dataflow Architecture}\label{sec:hw_arch}

Inspired by FPGA-based accelerators for low-precision quantized neural networks with static dataflow architectures, such as those created using the HLS4ML~\cite{duarte2018fast, ngadiuba2020compressing} and FINN~\cite{umuroglu2017finn} frameworks, this work adopts a fully static dataflow architecture aimed at maximum throughput and minimal latency at the trade-off of resource usage. In this approach, all operations in the JEDI-linear model are fully unrolled and mapped directly to physical hardware units. Specifically, all modules of the network are first converted into a monolithic combinational logic block, then broken into separate pipeline stages, i.e., registers may be inserted inside a layer, or the connection between two layers may be unregistered. Because no resource sharing is in place, each operation has a dedicated execution path, and the architecture has a fixed initiation interval and deterministic latency with minimal control overhead.

Compared to the conventional single-engine design~\cite{fan2018real, que2021recurrent}, which reuses a common computation unit across all layers, the proposed approach allows each block in JEDI-linear to be independently optimized for resource utilization, numerical precision, and computational structure. As a result, it leverages the full potential of FPGA customizability and improves scalability when processing a jet. Data is kept on-chip throughout the pipeline to avoid repeated off-chip memory accesses, significantly reducing end-to-end inference latency.

\subsection{Automation}

To facilitate rapid and scalable hardware deployment of JEDI-linear, we develop a fully automated flow that converts high-level Python models into synthesizable RTL, leveraging and extending the symbolic tracing capabilities of the \texttt{da4ml} framework~\cite{sun2025da4ml}. This automation is important as JEDI-linear's fully unrolled architecture for ultra-low latency cannot be efficiently handled by existing HLS compilers due to large loop bounds and structural complexity.

For each layer or operator type present in the JEDI-linear model, we design corresponding symbolic functions compatible with da4ml's abstraction. The model is then ``replayed'' over symbolic inputs, mirroring its numerical execution and producing a symbolic computation graph. During this process, high-level neural operations are decomposed into low-level arithmetic primitives, such as bitwise logic and shift-add trees, which are amenable to fully-unrolled, pipelined hardware implementations. Our extended version of \texttt{da4ml} then applies a custom pipelining pass to generate register-balanced pipelines for each subgraph, ensuring deterministic latency and initiation interval across all model configurations. The resulting computation graph is translated into synthesizable Verilog, supporting seamless hardware synthesis without manual intervention. The generated RTL models are functionally validated using Verilator~\cite{verilator}, with all tested configurations showing bit-exact results against their original quantized Python models.

This approach not only enables push-button deployment of JEDI-linear variants but also serves as a general framework for automatically translating quantized, unrolled neural networks into efficient FPGA implementations—paving the way for broader adoption in real-time scientific computing environments.

\section{Evaluation and Analysis}

\subsection{Experimental Setup}

The hls4ml jet tagging dataset~\cite{hls4ml-dataset, dataset2} is a common benchmark dataset targeting real-time event processing widely adopted by the high energy physics community~\cite{moreno2020jedi, que2022opt,  que2024ll, odagiu2024ultrafast, sun2025fast, nas-fpga}.
Five classes of jets, categorized by their originating particles, are included: gluons (g), light quarks (q), W bosons (W), Z bosons (Z), and top quarks (t). This dataset has 620,000 jets in the training set and 260,000 jets in the test set, both balanced between the different classes. The number of particles per jet varies, with a maximum of 150 particles per jet. Each particle has 16 kinematic features~\cite{moreno2020jedi}. Python 3.11, Keras 3.10 with Jax 0.7.0, and CUDA 12.9 are used for training.

As shown in \cite{sun2025fast}, breaking permutation invariance with heterogeneous quantization on the particle dimension could lead to additional resource savings. However, as there may be scenarios where full-sorting the input particles is not feasible, we consider both the non-permutation-invariant and permutation-invariant cases. We consider the cases where the maximum number of input particles is $8, 16, 32, 64,$ and $128$.

Following \cite{sun2025fast,que2024ll, odagiu2024ultrafast}, we also consider two sets of input features: (1) the full set of 16 features, and (2) a reduced set of 3 features, where only $p_T$, $\eta$, and $\phi$ are used, with particles with $p_T < 2$ GeV removed to model the effect of upstream selection. With all combinations stated above, we have a total of 20 different configurations for the JEDI-linear model. For each configuration, we run a Pareto optimization to find the models with the best trade-off between accuracy and resource usage estimated by EBOPs by gradually increasing the penalty on large EBOPs in a single training run. All models on the Pareto frontier defined by EBOPs and validation accuracy are then implemented on FPGAs using the da4ml framework~\cite{sun2025da4ml} via Verilog code generation, where we pipeline every 2 adders for an $F_{max}$ of approximately 300 MHz. All accuracies reported are based on RTL simulation with Verilator 5.034 of the generated firmware on the test set, and all hardware metrics are obtained after place and routing with Vivado 2025.1. The AMD VU13P FPGA is used in all evaluations.

\begin{figure}
   \begin{center}
      \includegraphics[width=0.8\linewidth]{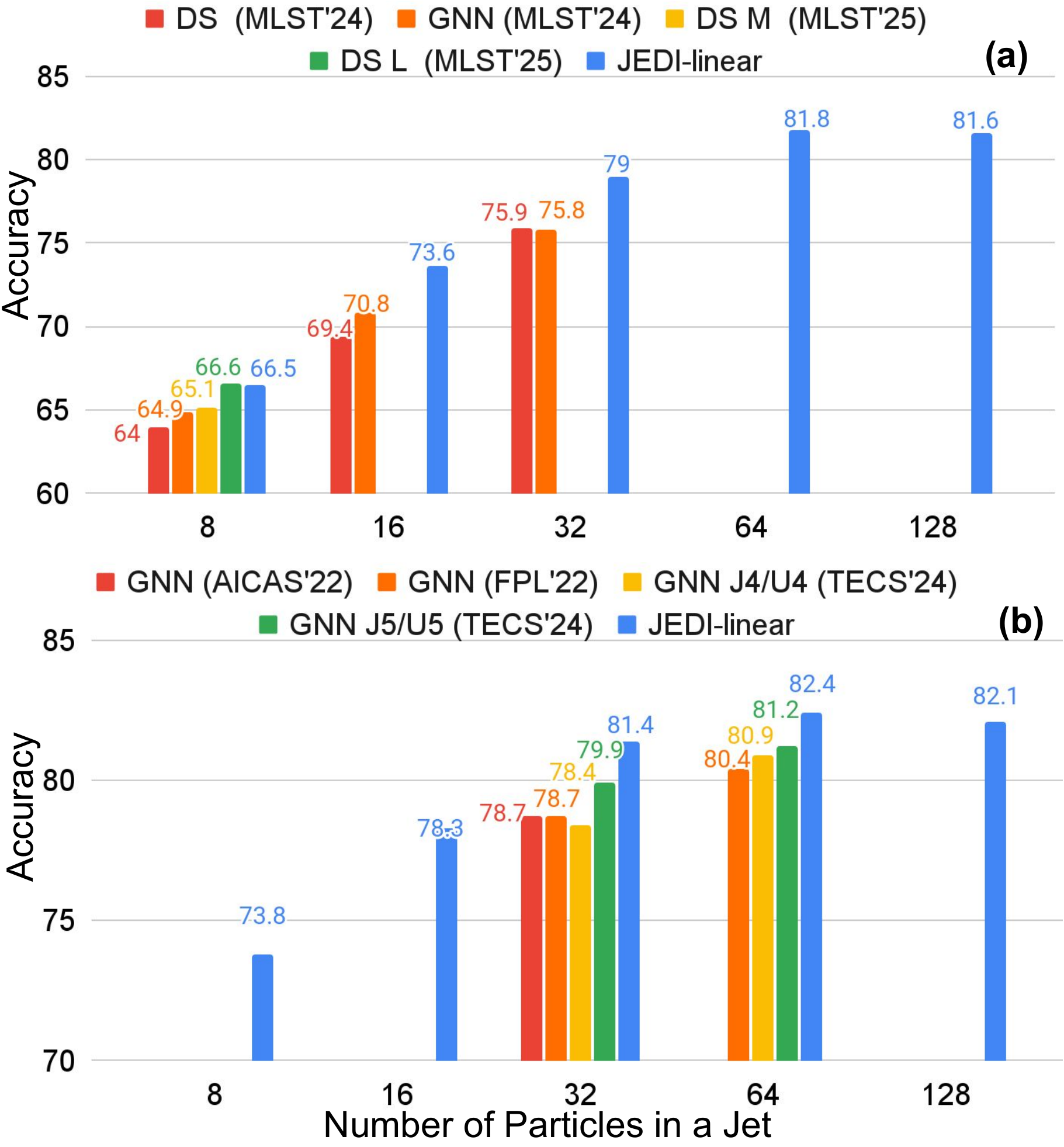}
   \end{center}
   \vspace{-0.3cm}
   \caption{Model accuracy across different numbers of particles per jet. (a) Comparison with prior DS and GNN-based methods using 3-feature inputs, including Ultrafast DS and GNN~\cite{odagiu2024ultrafast} and DS M/L~\cite{ds-fpga}. (b) Comparison with JEDI-net variants~\cite{que2022reconf, que2022opt, que2024ll} using 16-feature inputs.}
   \label{fig:acc}
\end{figure}

\begin{figure}
   \begin{center}
      \vspace{-0.3cm}
      \includegraphics[width=0.8\linewidth]{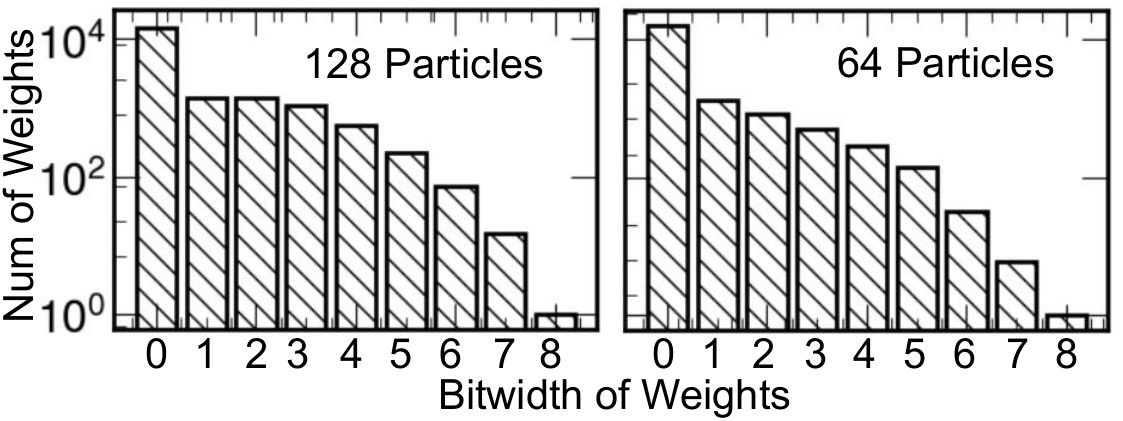}
   \end{center}
   \vspace{-0.3cm}
   \caption{Bitwidth of trained JEDI-linear models with datasets of 64/128 particles and 16 features.}
   \label{fig:bitwidth}
\end{figure}

\subsection{Model Accuracy}
Fig.~\ref{fig:acc} presents the model accuracy under two different input feature settings. In Fig.~\ref{fig:acc}(a), where only 3 features are used, our method outperforms previous approaches including Ultrafast DeepSets (DS) and JEDI-net (GNN)~\cite{odagiu2024ultrafast}, and DS M/L~\cite{ds-fpga} by significant margins that increase as the number of particles increases. Fig.~\ref{fig:acc}(b) illustrates the comparison between more expressive models using all 16 features and the GNN designs proposed in references~\cite{que2022reconf, que2022opt, que2024ll}. It is shown that JEDI-linear not only achieves higher accuracy across all tested particle counts, but also exhibits better scalability and stability when handling larger input sequence lengths. Unlike prior designs that rely on uniform bitwidth quantization, our model employs fine-grained mixed-precision quantization for more efficient model compression. We show the resulting weight bitwidth distributions in Fig.~\ref{fig:bitwidth} to illustrate the effectiveness of our quantization-aware training. It can be seen that the final weights are highly sparse, and the majority of non-zero weights have less than 3 bits, all while maintaining high model accuracies, illustrating the effectiveness of our quantization-aware training method.

\begin{table*}
      \centering

      \newcommand{\pminv}{\begin{tabular}[c]{@{}c@{}}Perm. \\ Inv. \end{tabular}}
      \newcommand{\acc}{\begin{tabular}[c]{@{}c@{}}Acc. \\ (\%) \end{tabular}}
      \newcommand{\lat}{\begin{tabular}[c]{@{}c@{}}Latn. \\ (ns) \end{tabular}}
      \newcommand{\lut}{\begin{tabular}[c]{@{}c@{}}LUT \\ (k) \end{tabular}}
      \newcommand{\ff}{\begin{tabular}[c]{@{}c@{}}FF \\ (k) \end{tabular}}
      \newcommand{\fmax}{\begin{tabular}[c]{@{}c@{}} $F_\mathrm{max}$ \\ (MHz) \end{tabular}}

      \caption{Comparison of permutation-invariant JEDI-linear models and other state-of-the-art models with \textbf{3/16 input features} on AMD VU13P FPGAs. For these models, the top-$N$ particles are selected based on $p_T$ for the input.}
      \label{tab:compare-gnn}
      \scalebox{1.08}{\begin{tabular}{c|ccccccccccc}
                  \toprule
                  Model                               & Particles    & Features & \acc    & \lat  & DSP    & \lut  & \ff & BRAM & II (clk) & \fmax
                  \\

                  \midrule
                  DS (MLST'24)~\cite{odagiu2024ultrafast}         & 8            & 3       & $<64.0$ & 95    & 626    & 386   & 121 & 4    & 3        & N/A   \\
                  DS (MLST'24)~\cite{odagiu2024ultrafast}         & 16           & 3       & $<69.4$ & 115   & 555    & 747   & 239 & 4    & 3        & N/A   \\
                  DS (MLST'24)~\cite{odagiu2024ultrafast}         & 32           & 3       & $<75.9$ & 130   & 434    & 903   & 359 & 4    & 2        & N/A   \\
                  \midrule
                  GNN (MLST'24)~\cite{odagiu2024ultrafast}        & 8            & 3       & $<64.9$ & 160   & 2,120  & 472   & 192 & 132  & 3        & N/A   \\
                  GNN (MLST'24)~\cite{odagiu2024ultrafast}        & 16           & 3       & $<70.8$ & 180   & 5,362  & 1,388 & 594 & 52   & 3        & N/A   \\
                  GNN (MLST'24)~\cite{odagiu2024ultrafast}        & 32           & 3       & $<75.8$ & 205   & 2,120  & 1,162 & 761 & 12   & 3        & N/A   \\
                  \midrule
                  DS M (MLST'25)~\cite{nas-fpga}      & 8            & 3       & 65.1    & 110   & 548    & 130   & 49  & 4    & 3        & N/A   \\
                  DS L (MLST'25)~\cite{nas-fpga}      & 8            & 3       & 66.6    & 135   & 2,458  & 337   & 140 & 4    & 3        & N/A   \\
                  \midrule
                  \textbf{JEDI-linear}                  & 8            & 3       & 66.5    & 79    & 0      & 136   & 73  & 0    & 1        & 302.8 \\
                  \textbf{JEDI-linear}                  & 16           & 3       & 73.6    & 75    & 0      & 136   & 71  & 0    & 1        & 305.7 \\
                  \textbf{JEDI-linear}                  & 32           & 3       & 79.0    & 80    & 0      & 136   & 79  & 0    & 1        & 299.4 \\
                  \textbf{JEDI-linear}                  & \textbf{64}  & 3       & 81.8    & 78    & 0      & 164   & 93  & 0    & 1        & 307.0 \\
                  \textbf{JEDI-linear}                  & \textbf{128} & 3       & 81.6    & 138   & 0      & 296   & 163 & 0    & 1        & 203.1 \\
                  \midrule
                  \midrule
                  GNN (AICAS'22)~\cite{que2022reconf} & 30           & 16      & 78.7    & 3000  & 7417   & 810   & 205 & 924  & 600      & N/A   \\
                  \midrule
                  GNN (FPL'22)~\cite{que2022opt}      & 30           & 16      & 78.7    & 1910  & 11504  & 1158  & 246 & 1392 & 400      & N/A   \\
                  GNN (FPL'22)~\cite{que2022opt}      & 50           & 16      & 80.4    & 10660 & 12,284 & 1515  & 533 & 1607 & 650      & N/A   \\
                  \midrule
                  GNN J4 (TECS'24)~\cite{que2024ll}   & 30           & 16      & 78.4    & 290   & 8,776  & 865   & 138 & 37   & 30       & N/A   \\
                  GNN J5 (TECS'24)~\cite{que2024ll}   & 30           & 16      & 79.9    & 905   & 9,833  & 911   & 158 & 37   & 150      & N/A   \\
                  GNN U4 (TECS'24)~\cite{que2024ll}   & 50           & 16      & 80.9    & 650   & 8,945  & 855   & 201 & 25   & 100      & N/A   \\
                  GNN U5 (TECS'24)~\cite{que2024ll}   & 50           & 16      & 81.2    & 905   & 8,986  & 815   & 189 & 37   & 150      & N/A   \\
                  \midrule
                  \textbf{JEDI-linear}                  & 8            & 16      & 73.8    & 67    & 0      & 72    & 40  & 0    & 1        & 311.3 \\
                  \textbf{JEDI-linear}                  & 16           & 16      & 78.3    & 72    & 0      & 99    & 50  & 0    & 1        & 307.0 \\
                  \textbf{JEDI-linear}                  & 32           & 16      & 81.4    & 79    & 0      & 147   & 71  & 0    & 1        & 304.7 \\
                  \textbf{JEDI-linear}                  & 64           & 16      & 82.4    & 93    & 0      & 192   & 92  & 0    & 1        & 268.1 \\
                  \textbf{JEDI-linear}                  & \textbf{128} & 16      & 82.1    & 110   & 0      & 243   & 111 & 0    & 1        & 237.4 \\
                  \midrule
            \end{tabular}}
            \vspace{-0.2cm}
\end{table*}

\subsection{Performance Analysis and Discussion}

Table~\ref{tab:compare-gnn} presents a comprehensive comparison between our proposed quantized JEDI-linear models and a variety of state-of-the-art models, focusing on permutation-invariant architectures with a fixed number of input features (3 or 16). The models are evaluated in terms of classification accuracy, latency, and hardware resource utilization (DSP, LUT, FF, BRAM, initiation interval (II), and maximum clock frequency).

Compared to existing designs, our JEDI-linear models demonstrate a significant improvement in latency and initiation interval across all configurations. For example, with 16 input features and 64 particles, our design achieves an accuracy of 81.4\% at only 79~ns latency, outperforming models such as GNN U4 and U5~\cite{que2024ll} and GNN (FPL'22)~\cite{que2022opt} which require latencies of 905 ns and 10,660 ns, respectively, for higher accuracies. 
Table~\ref{tab:compare-gnn} also shows that JEDI-linear achieves lower latency across various jet sizes compared to prior designs.
Even at the largest configuration with 128 particles, our latency remains as low as 110~ns, which is suitable for hardware trigger systems that demand ultra-fast inference.
In addition, our designs achieve a one-cycle initiation interval, a feat never before achieved by previous GNN designs.

\begin{table*}
      \centering

      \newcommand{\pminv}{\begin{tabular}[c]{@{}c@{}}Perm. \\ Inv. \end{tabular}}
      \newcommand{\acc}{\begin{tabular}[c]{@{}c@{}}Acc. \\ (\%) \end{tabular}}
      \newcommand{\lat}{\begin{tabular}[c]{@{}c@{}}Latn. \\ (ns) \end{tabular}}
      \newcommand{\lut}{\begin{tabular}[c]{@{}c@{}}LUT \\ (k) \end{tabular}}
      \newcommand{\ff}{\begin{tabular}[c]{@{}c@{}}FF \\ (k) \end{tabular}}
      \newcommand{\fmax}{\begin{tabular}[c]{@{}c@{}} $F_\mathrm{max}$ \\ (MHz) \end{tabular}}

      \caption{Comparison of JEDI-linear non-permutation-invariant models with state-of-the-art models with \textbf{3/16 input features} on AMD VU13P FPGAs. For these models, the input particles are sorted by $p_T$ before feeding into the model. }
      \label{tab:compare-mlpm}
      \scalebox{1.08}{\begin{tabular}{c|ccccccccccc}
                  \toprule
                  Model                             & Particles & Features & \acc & \lat & DSP & \lut & \ff & BRAM & II (clk) & \fmax
                  \\
                  \midrule
                  MLPM (MLST'25)~\cite{sun2025fast} & 16        & 3        & 71.7 & 68   & 0   & 75   & 17  & 0    & 1        & 205.5 \\
                  MLPM (MLST'25)~\cite{sun2025fast} & 32        & 3        & 78.0 & 62   & 0   & 63   & 15  & 0    & 1        & 211.0 \\
                  MLPM (MLST'25)~\cite{sun2025fast} & 64        & 3        & 79.7 & 72   & 0   & 159  & 36  & 0    & 1        & 209.4 \\
                  MLPM (MLST'25)~\cite{sun2025fast} & 128       & 3        & 79.8 & 72   & 0   & 83   & 21  & 0    & 1        & 208.7 \\

                  \midrule

                  \textbf{JEDI-linear}                & 16        & 3        & 71.9 & 54   & 0   & 44   & 22  & 0    & 1        & 354.4 \\
                  \textbf{JEDI-linear}                & 32        & 3        & 78.0 & 63   & 0   & 45   & 26  & 0    & 1        & 300.8 \\
                  \textbf{JEDI-linear}                & 64        & 3        & 80.9 & 61   & 0   & 71   & 38  & 0    & 1        & 328.4 \\
                  \textbf{JEDI-linear}                & 128       & 3        & 80.9 & 82   & 0   & 98   & 48  & 0    & 1        & 257.6 \\

                  \midrule
                  MLPM (MLST'25)~\cite{sun2025fast} & 16        & 16       & 77.5 & 71   & 0   & 102  & 24  & 0    & 1        & 210.9 \\
                  MLPM (MLST'25)~\cite{sun2025fast} & 32        & 16       & 80.7 & 65   & 0   & 87   & 22  & 0    & 1        & 215.8 \\
                  MLPM (MLST'25)~\cite{sun2025fast} & 64        & 16       & 81.6 & 65   & 0   & 126  & 32  & 0    & 1        & 213.9 \\
                  MLPM (MLST'25)~\cite{sun2025fast} & 128       & 16       & 81.3 & 77   & 0   & 151  & 42  & 0    & 1        & 207.8 \\
                  \midrule

                  \textbf{JEDI-linear}                & 16        & 16       & 77.6 & 52   & 0   & 38   & 20  & 0    & 1        & 381.7 \\
                  \textbf{JEDI-linear}                & 32        & 16       & 80.9 & 60   & 0   & 62   & 33  & 0    & 1        & 347.7 \\
                  \textbf{JEDI-linear}                & 64        & 16       & 81.8 & 67   & 0   & 84   & 47  & 0    & 1        & 327.8 \\
                  \textbf{JEDI-linear}                & 128       & 16       & 81.7 & 77   & 0   & 93   & 46  & 0    & 1        & 285.7 \\
                  \midrule
            \end{tabular}}
            \vspace{-0.2cm}
\end{table*}

A major advantage of our approach lies in its efficient hardware with low resource utilization. Unlike prior works such as Ultrafast DS and GNN~\cite{odagiu2024ultrafast} which require hundreds to thousands of DSPs, our models use zero DSP blocks across all configurations. In order to achieve ultra-low latency, Ultrafast GNN~\cite{odagiu2024ultrafast} trades off as many hardware resources as possible by increasing the parallelism to achieve low latency. For instance, its model with 16-particle inputs that utilizes 1,388k (over 80\%) LUTs on VU13P FPGAs. It is hard to achieve low latency with a large input of more particles.
With our proposed efficient neural architecture and hardware-aware optimizations, our JEDI-linear model with 16-particle inputs not only utilizes 10.2 times fewer LUTs, but also achieves around 3\% higher model accuracy as well as 2.4 times lower latency and 3 times lower initiation interval.
In addition, the GNN J5 model~\cite{que2024ll} with 79.9\% accuracy uses 9,833 DSPs, while our 32-particle model reaches a higher accuracy of 81.4\% without any DSP usage and with 6.2 times lower LUT usage, and 11.5 times lower latency. This makes our design far more viable for deployment on resource-constrained FPGAs while achieving high performance.

Our work also demonstrates good scalability. As the number of input particles increases from 8 to 128, classification accuracy steadily improves, from 66.5\% to 81.8\% with 3 features, and from 73.8\% to 82.4\% with 16 features, as shown in Fig.~\ref{fig:acc} and Table.~\ref{tab:compare-gnn}, while maintaining a consistent initiation interval of 1 clock cycle.
Our models also achieve high maximum clock frequencies, enabling continuous data processing at a very high throughput, which is critical for real-time decision-making systems.

\subsection{Comparison with Non-Permutation Invariant Models}

We also compare our JEDI-linear models with the recent MLP-Mixers (MLPM) designs~\cite{sun2025fast}, which are not permutation-invariant and rely on sorting particles by transverse momentum ($p_T$). For these models, we enable particle-wise quantization, which breaks the permutation invariance of the model but brings further resource savings by allowing selective feature prioritization on different particles.

\begin{figure}
   \begin{center}
      \includegraphics[width=1.0\linewidth]{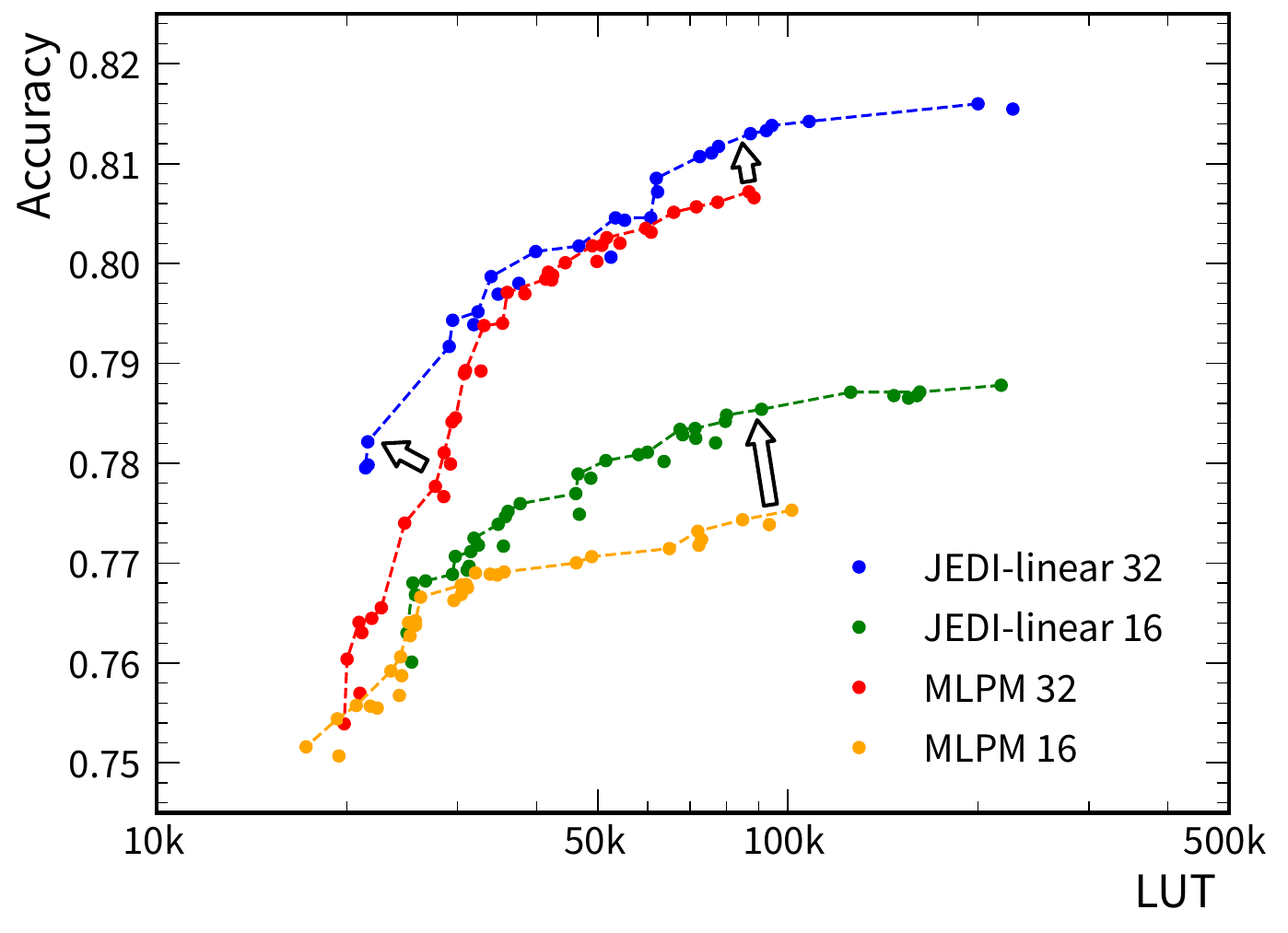}
   \end{center}
   \vspace{-0.5cm}
   \caption{Comparison of the JEDI-linear model with MLP-Mixer on the 16 feature dataset. Each point represents a model recorded during training on the Pareto Frontier defined by validation accuracy and EBOPs. The actual Pareto Frontier between test accuracy and LUT usage is shown in the dashed lines.}
   \label{fig:pareto}
\end{figure}

In Table~\ref{tab:compare-mlpm}, we present the non-permutation-invariant designs of JEDI-linear that have higher or equal accuracy compared to the MLP-Mixers from~\cite{sun2025fast}. At higher or equal accuracies, our designs achieve consistently lower resource utilization and latency under various configurations. For example, JEDI-linear achieves 77.6\% accuracy at just 52~ns with 16 particles and 16 features with 38k LUTs, compared to MLPM's 71~ns with 102k LUTs, resulting in 1.3 times lower latency and 2.7 times fewer LUTs. In addition, our models run at higher clock frequencies (up to 381.7~MHz vs. 215.8~MHz). The latency could be further reduced if the required clock rate is lower is a less aggressive pipelining strategy is be used. The Pareto plot in Fig.~\ref{fig:pareto} further confirms our advantage, showing that our models dominate MLPM in the accuracy-LUT trade-off. Overall, our JEDI-linear models achieve superior performance with better hardware efficiency, establishing a new Pareto frontier for FPGA-based jet tagging.

\section{Conclusions and Future Work}

This work introduces JEDI-linear, a novel GNN architecture that enables linear-complexity and low-latency jet tagging on FPGAs.
By eliminating explicit pairwise interactions and applying fine-grained quantization and distributed arithmetic, our designs achieve up to 82.4\% classification accuracy, reduce inference latency to below 60~ns, and eliminate DSP usage entirely.
Our work demonstrates that with careful algorithm-hardware co-design, high-performance GNNs can be deployed in real-time systems. In the future, we plan to explore broader GNN variants, automation using metaprogramming (e.g., MetaML~\cite{que2023metaml, que2025metaml}) and extend JEDI-linear to other scientific domains.

\noindent \textbf{Acknowledgement.}
Partial support from the United Kingdom EPSRC (grant numbers UKRI256, EP/V028251/1, EP/N031768/1, EP/S030069/1, and EP/X036006/1), and STFC (grant numbers ST/Y509115/1, ST/R005788/1, ST/W000490/1, ST/Z000149/1, ST/W000636/1, and ST/R005745/1), and United States DOE (grant number DE-SC0011925), and NSF ACCESS (grant number PHY240298), Intel and Lenovo (ICICLE HPC \& AI partnership) and AMD is gratefully acknowledged.

\footnotesize
\balance
\bibliographystyle{IEEEtran}
\bibliography{main-bibliography}

\end{document}